\documentclass[twoside]{article}

\usepackage[pdftex]{graphicx} %LaTeX2e

\usepackage{graphics}
\newcommand{\sip}[0]{\rightarrow}%

\begin{document}

\begin{center}
{\Large\sf{\noindent  Mathematical Model of DSB Formation by Ionizing Radiation}}
\vspace{0.5cm}

J. Barilla*, M. Lokaj\'{i}\v{c}ek**, P. Simr*

\vspace{0.2cm}
*J. E. Purkinje University in Usti nad Labem, Faculty of Science

**Institute of Physics, Academy of Sciences, Prague, Czech Republic
\end{center}

\vspace{0.5cm}

{\bf\sf\noindent }
%\vspace{0.5cm}
 {\large\sf  Abstract}

The understanding of inactivation radiobiological mechanism in individual cells is important when from one side the application of ionizing radiation to tumour therapy and from the other side the protection against radiation are to be effectively improved. Important part of this mechanism is double-strand break (DSB) formation; these DSB may be formed directly by impacting ionizing particles or indirectly by different secondary radicals. The latter kind of formation is much more frequent when cells contain normal water content. Mathematical model of the corresponding chemical stage will be presented with the aim to demonstrate how the individual radicals, but also other chemical agents present in the water during irradiation may influence DSB formation.
\\   


\section*{\large\sf Introduction}
	
Ionizing radiation represents one of the main methods of tumour treatment. The aim is to destroy tumour focus without damaging irreversibly surrounding tissues. To fulfill this goal it is necessary to understand well the radiobiological processes running in individual living cells after the impact of a ionizing particle. The given results will be helpful also as to radiation protection.

According to contemporary knowledge the radiobiological processes in a normal cell (with usual water content)  may be divided into three phases: physical (energy transfer from radiation to medium and the formation of water radicals), chemical (reactions initiated by radicals and formation of damages in chromosomal DNA) and biological (reaction of individual cells to the damage ending or by damage repair or by cell inactivation).  The first two phases are finished in a very short time interval (fraction of second) so that at dose-rates applied to in radiotherapy treatment (or playing role in radiation protection) the effects of individual ionizing particles occur practically quite independently. In the radiotherapy the total dose is delivered in a rather short time so that the total DNA damage gathered during one irradiation may be regarded as the starting point of the biological phase ending after much longer time (at least ten minutes or longer).

It is possible to say that the final result of biological phase depends on the number and distribution of individual DSB (double strand breaks) in the system of chromosomes. Thus the probability of DSB formation by one ionizing particle (or photon) may be taken as the basic parameter determining inactivation effect. In the following we will try to analyze the importance of individual processes running in the chemical phase with the help of a corresponding mathematical model.

However, it is necessary to distinguish between the ionizing particles exhibiting different ionization densities. Photons, accelerated electrons, and in principle also protons belong to low-density or low-LET (linear energy transfer) radiation kinds when the cell being hit by one particle only are not inactivated as a rule. On the other side, the single accelerated heavier atom nuclei (high-LET radiation) inactivate cells having be hit by their Bragg peaks. As to the formation of DSB by low-LET radiation it may be rather strongly influenced by the presence of other (radiosensitive or radioprotective) chemical agents. And only rather sophisticated (complex) mathematical models may be helpful in understanding better corresponding mechanism.

\section*{\large\sf Chemical phase and DSB formation}
    
We have already mentioned that the inactivation effect is approximately proportional to the efficiency of DSB formation. However, it depends also on the distribution of DSB in chromosomal system, which may be very  different at a given dose for divers radiation types. The efficiency may be studied by measuring the survival curves, i.e. the ratio of surviving (not inactivated) cells for different radiation doses. And it is also possible to establish experimentally the numbers of DSB in the given chromosomal DNA under the same conditions. The latter method enables then to establish some important characteristics also outside the biological material for corresponding DNA molecules dissolved in water.

One may assume that in DNA dissolved in water (i.e., also in normally living cells) practically all SSB and DSB are formed in indirect way by radicals formed in water. A direct collision of an individual ionizing particle with a given DNA may by practically neglected. The radical clusters of different sizes are formed along particle tracks. However, the most of them are so small that only one SSB may be formed with a certain probability. As to the low-LET radiation they are only clusters formed by track ends of electrons of greater original energy that may form a DSB with sufficient probability. Only heavier ionizing particles in their Bragg peaks are more efficient. The conclusion that DSB is practically always formed by a single radical cluster is strongly supported by the fact that the number of DSB increases linearly with increasing dose \cite{frank}. 

And we should ask what is the size of clusters that may form a DSB and how is the DSB formation influenced by chemical mechanism in individual clusters.
Our goal is to study the reactions running in radical clusters and to analyze how their final effect depends on their size and might be influenced by the presence of other chemical agents; and what is the average size of radicals forming DSB for low-LET radiation. The problem may be hardly solved without the help of suitable mathematical models and computer simulations of the corresponding processes.

After their impact  the ionizing particles transfer their energy to the water medium and form clusters of water radicals (eventually, also of other ones) when other corresponding chemical agents are present, too. Oxygen radicals are usually involved under standard aerobic conditions. And it is known that the radiobiological effect increases with oxygen content, which should be accompanied by the corresponding increase of SSB and DSB numbers. It is important in radiotherapy as for low-LET radiation the effect in anaerobic tumor cells diminishes. 

Irradiating the cells with the normal content of water the chemical stage of radiobiological mechanism may have significant influence on the final effect. The following processes are then running in the corresponding radical clusters:
\begin{enumerate}
	\item[(i)] the free radicals having been formed by ionizing particle may mutually react and may recombine and form new chemical substances that react further;
 	\item[(ii)] the radical clusters diffuse into surrounding medium, the density of radicals diminishes and their number gradually decreases;
	\item[(iii)] all substances present in the cluster (not only radicals) may take part in chemical reactions and other agents (including radicals) may be formed;
 	\item[(iv)] different agents (mainly chemically active radicals) present in the cluster may react with cell DNA and the biologically efficient DSB  in chromosomal DNA may be formed.
\end{enumerate}
And it is evident that the efficiency of DSB formation will depend not only on the size of individual clusters, but also on all chemical agents present in them during irradiation, important role being played also by the oxygen in aerobic conditions. It is also the diffusion of individual clusters that may play an important role in final effect. And it is practically evident that rather complex mathematical models are needed when we are to understand the efficiency of individual radicals in SSB (and DSB) formation. 

We have proposed a corresponding mathematical model involving the influence of chemical reactions as well as of diffusion; its preliminary version having been presented in Ref. \cite{baril}. We have demonstrated its application using the experimental data obtained by Block and Loman \cite{blok} in irradiating DNA molecules of bacteriofague $\Phi$X174 dissolved in water by Co gamma radiation (of energy cca 1,25MeV). The frequency of SSB (and DSB) formation in dependence on oxygen content in the mixture with $N_2$ and $N_2O$  were determined in this experiment.

\section*{\large\sf Processes running in chemical phase}

The chemical processes running in water after the impact of ionizing particles has been described e.g. in the paper of Sauer and Schmidt \cite{sauer}. Here a short survey of them will be given and their role in DSB formation will be mentioned.

One may expect that they are mainly water radicals of $H^{\bullet}$, $OH^{\bullet}$ and $e^{-}_{aq}$ that form individual SSB, and also $HO^{\bullet}_{2}$ radical if oxygen is present.
 And one may ask what is the average size of the radical clusters that would be consistent with frequencies of DSB established experimentally under different conditions.
 While $HO^{\bullet}_{2}$ radical is formed in secondary reactions with oxygen all water radicals are formed during the impact of ionizing particles. First of all it is the ionization of water molecule (ionization potential 12,56 ev):
\begin{equation}
   H_{2}O \rightarrow H_{2}O^{+} + e^{-};
\label{eq1}
\end{equation}
 $H_{2}O^{+}$ ion may decay and produce $OH{^\bullet}$ radicals:
\begin{equation}
H_{2}O^{+} \rightarrow H^{+} + OH^{\bullet}.
\label{eq2}
\end{equation}
If the energy of a photon is approximately 7 eV the water molecule is excited and the following dissociation  leads again
to the formation of water radicals:
\begin{equation}
   H_{2}O \rightarrow H_{2}O^{*} \rightarrow H^{\bullet} + OH^{\bullet}.
\label{eq3}   
\end{equation}

\begin{table}[htbp]
\centering
\small
\begin{tabular}{|rrcl|r|}
\hline
\multicolumn{4}{|c|}{\bf Reactions}&{\bf Rate constants}\\
&&&&(dm$^{3}\cdot $mol$^{\mathrm -1}\cdot$s$^{-1})$\\
\hline
1.&$ H^\bullet + H^\bullet $&$\sip$&$ H_{2}  $&$1 \cdot 10^{10}$\\
2.&$ e_{aq}^{-} + H^\bullet $&$\sip$&$ H_{2} + OH^{-} $&$2.5
\cdot 10^{10}$\\
3.&$ e_{aq}^{-} + e_{aq}^{-} $&$\sip$&$ H_{2} + 2OH^{-}$&$6
\cdot 10^{10}$\\
4.&$ e_{aq}^{-} + OH^\bullet $&$\sip$&$ OH^{-} + H_{2} O $&$3
\cdot 10^{10}$\\
5.&$ H^\bullet + OH^\bullet $&$\sip$&$ H_{2} O  $&$2.4 \cdot 10^{10}$\\
6.&$ OH^\bullet + OH^\bullet $&$\sip$&$ H_{2} O_{2}  $&$4 \cdot 10^{9}$\\
7.&$ H_{3} O^{+} + e_{aq}^{-} $&$\sip$&$ H^\bullet + H_{2} O$&$2.3
\cdot 10^{10}$\\
8.&$ H_{3} O^{+} + OH^{-} $&$\sip$&$ H_{2} O$&$3 \cdot 10^{10}$\\
9.&$ H^\bullet + H_{2} O_{2} $&$\sip$&$ H_{2} O + OH^\bullet $&$1
\cdot 10^{8}$\\
10.&$ e_{aq}^{-} + H_{2} O_{2} $&$\sip$&$ OH^\bullet + OH^{-}$&$1.2
\cdot 10^{10}$\\
11.&$ OH^\bullet + H_{2} O_{2} $&$\sip$&$ H_{2} O + HO_{2}^\bullet$&$5
\cdot 10^{7}$\\
12.&$ OH^\bullet + H_{2} $&$\sip$&$ H_{2} O + H^\bullet $&$6 \cdot 10^{7}$\\
13.&$ HO_{2}^\bullet + H^\bullet $&$\sip$&$ H_{2} O_{2}  $&$1
\cdot 10^{10}$\\
14.&$ e_{aq}^{-} + O_{2} $&$\sip$&$ O_{2}^{-} + H_{2} O$&$1.9
\cdot 10^{10}$\\
15.&$ HO_{2}^\bullet + OH^\bullet $&$\sip$&$ H_{2} O + O_{2}$&$1
\cdot 10^{10}$\\
16.&$ HO_{2}^\bullet + HO_{2}^\bullet $&$\sip$&$ H_{2} O_{2} +O_{2}
$&$2 \cdot 10^{6}$\\
17.&$ H^\bullet + O_{2} $&$\sip$&$ HO_{2}^\bullet  $&$1 \cdot 10^{10}$\\
18.&$ O_{2}^{-} + H_{3} O^{+} $&$\sip$&$ HO_{2}^\bullet$&$3
\cdot 10^{10}$\\
19.&$ H_{2} O $&$\sip$&$ H_{3} O^{+} + OH^{-}$&$5.5 \cdot 10^{-6}$\\
20.&$ HO_{2}^\bullet $&$\sip$&$ H_{3} O^{+} + O_{2}^{-}$&$1 \cdot 10^{6}$\\
\hline
\end{tabular}
\caption{Recombination reactions going in water after radiation impact and their rate constants \cite{chat}} 
\label{tab:1}
\end{table}

The electron released in reaction (\ref{eq1}) may bind to a water molecule
\begin{equation}
e^{-} + (H_{2}O)_{n} \rightarrow e^{-}_{aq}
\label{eq4}
\end{equation}
and electron $e^-_{aq}$  contributes to production of hydrogen radicals:
\begin{equation}
e^{-}_{aq} + H_{2}O \rightarrow OH^{-} + H^{\bullet}.
\label{eq5}
\end{equation}
Hydrated electrons $e^-_{aq}$ may exist in this form relatively long; they may diffuse to greater distances and react with DNA molecules. 

The effect of ionizing radiation is known to be modified if  oxygen is present in the solution. It represents a source of some radicals which are assumed to be aggressive. They may arise by the reaction
\begin{equation}
e^{-} + O_{2} \rightarrow O^{\bullet} + O^{-};
\label{eq6}
\end{equation}
minimum electron energy $\sim$ 4 eV, maximum gain at 8 eV.
These radicals react in water medium and new radicals are formed:
\begin{eqnarray}
   O^{\bullet} + H_{2}O &\rightarrow& H^{\bullet} + HO^{\bullet}_{2},\label{eq7} \\
   O^{-} + H_{2} O &\rightarrow& H^{\bullet} + HO^{-}_{2}\label{eq8}.
\end{eqnarray}

Oxygen radicals may be involved or formed further in the reactions
\begin{eqnarray}
  e^{-}_{aq} + O_{2} &\rightarrow&  O^{-}_{2},\label{eq9} \\
  O^{-}_{2} + H_{3}O^{+} &\rightarrow&  HO^{\bullet}_{2},\label{eq10} \\
  H^{\bullet} + O_{2} &\rightarrow&  HO^{\bullet}_{2}, \label{eq11}\\
  HO^{\bullet}_{2} &\rightarrow& H_{3}O^{+} + O^{-}_{2}.\label{eq12}
\end{eqnarray}
At lower concentrations of $O_2$  (at pH7) the dissociation (\ref{eq12}) is preferred; at higher concentrations it is the formation of radicals  $HO^{\bullet}_{2}$ according to Eq. (\ref{eq10}) which is preferred \cite{pika}.

Thus, the actual content of radicals in an individual cluster depends also on oxygen concentration; the radicals $HO^{^\bullet}_{2}$ (resp. $O{^\bullet}$) may be involved in DSB formation in addition to $H{^\bullet}$ and $OH{^\bullet}$. Possible recombination processes and other reactions are summarized in Table \ref{tab:1}.

\section*{\large\sf Mathematical modeling of the chemical mechanism}

As already mentioned the mathematical model proposed by us earlier (see \cite{baril}) has been applied to fitting the experimentally determined numbers of DSB for different concentrations  of molecular oxygen in the mixtures of $O_{2}$ with $N_{2}$ and  $N_{2}O$  in water solution of  DNA. The proposed model was very simple. And we have succeeded in reproducing the data when the reaction rates (and also their mutual ratios) of involved chemical species differed rather significantly from those shown in Table 1. 

In the following we shall apply the model only to one set of experimental data (for the mixture of $O_2$ with $N_2$ only), but we will broaden its basis. First of all, the model will be partially generalized and the number of recombination reactions will be extended. However, on the other side some limiting conditions will be imposed on the parameters that were left earlier quite free:

 - the diffusion coefficients for individual radicals will be taken from the literature and fixed (see Table 2); 
 
 - it will be required for the ratios of the most reaction rates to correspond to experimentally established values (i.e., for mutual reactions of substances in clusters as well as for reaction of radicals with DNA); 

 - and also for original number of radicals in corresponding cluster it will be required to correspond to efficiency of reactions responsible for their formation. \\ 
Only several parameters (as will be mentioned later) will be left quite free.

\begin{table}[htbp]
	\centering
  \small
		\begin{tabular}{|cc|c|c|c|}
			\hline
			& & Diffusion coefficient & Number & Designation of\\
			\multicolumn{2}{|c|}{Substance} & $\left(nm^{2}.ns^{-1}\right)$ & of species & dif. coefficients\\
			\hline
			1. & $H^{\bullet}$ & 7.0 & $N_{H}$ & $D_{H}$\\
			2. & $OH^{\bullet}$ & 2.8 & $N_{OH}$ & $D_{OH}$\\
			3. & $e^{-}_{aq}$ & 4.5 & $N_{e}$ & $D_{e}$\\
			4. & $HO^{\bullet}_{2}$ & 2.0 & $N_{O}$ & $D_{O}$\\
		  5. & $H_{3}O^{+}$ & 9.0 & $N_{H_{3}O}$ & $D_{H_{3}O}$\\
			6. & $O^{-}_{2}$ & 2.1 & $N_{O^{-}_{2}}$ & $D_{O^{-}_{2}}$\\
			\hline
		\end{tabular}
	\caption{Diffusion coefficients}
	\label{tab:2}
\end{table}

The  mathematical model of corresponding chemical mechanism will start from the assumption of the existence of a hypothetical average cluster system characterized by nonhomogeneous concentrations of individual species (see \cite{Mozum}). Macroscopic laws will be used to describe the diffusion of radiation-induced objects and the concentration changes due to different chemical reactions. Such a situation may be described by the following set of coupled differential equations: 
\begin{equation}
\frac{\partial c_{i}}{\partial t}=D_{i}\nabla^{2}c_{i} - c_{i} \sum_{j}k_{ij}c_{j} +  \sum_{j,k\neq i}k^{(i)}_{jk}c_{j}c_{k},
\label{q3.6a}
\end{equation}
where $D_{i}$ are diffusion coefficients and $c_{i}$ - spatially dependent concentrations of species $i$; $k_{ij}$ are rate constant of reactions between species  $i$ and $j$. The first term of the right-hand side of the equation represent the diffusive contribution to the evolution of  $c_{i}$, while the second and third terms represent removals and productions of the $i$ substances by chemical reactions. The set of such coupled equations may be solved either by analytic approximation or numerically.

We will  consider the kinetics of low-LET radiolysis. The system will be described in terms of a spherically symmetric typical spur (cluster). We can then substitute equations (\ref{q3.6a}) by equations, where diffusion processes are considered as spherically symmetrical
\begin{equation}
   \frac{\partial c_{i}}{\partial t} = D_{i}\frac{1}{r^{2}} \frac{\partial}{\partial r}
\left( r^{2} \frac{\partial c_{i}}{\partial r} \right) - c_{i} \sum_{j}k_{ij}c_{j} +  \sum_{j,k\neq i}k^{(i)}_{jk}c_{j}c_{k} ;
\label{qB2}
\end{equation}
$r$ denoting the distance from the cluster center.

The most frequently used analytic models for the fast kinetics in radiolysis are based on  approximation suggested by Jaffe (see \cite{Jaffe}); they are known as prescribed diffusion. The initial spatial distributions of the radiation-induced particles, their concentration profiles, are assumed to be Gaussian, and the kinetic analysis invokes the approximation that reactions affect only the numbers of particles and not the form of the nonhomogeneous spatial profiles, which are therefore always Gaussian.

One could solve the given system of partial differential equations numerically but it would be rather time demanding to determine the parameters of mathematical model with the help of an optimization procedure (see e.g. \cite{james}) to fit the used experimental data. Therefore, we have made use of the method which enables to substitute the  solving of the system of partial differential equations by the solving of the system of ordinary differential equations, which is much  less time consuming.   

The concentration profiles of species i due to diffusion process only may be obtained by solving Eq. (\ref{qB2}) containing only the first term on the right-hand side; one obtains 
\begin{equation}
c_{i}(r,t) = \frac{N_{i}}{8\sqrt{\left(\pi D_{i}t\right)^{3}}} exp \left( - \frac{r^{2}}{4D_{i}t} \right)
\label{q1.37}
\end{equation}
where $N_{i}$ is the (initial) number of species $i$ in the cluster. This concentration dependence corresponds to the Gaussian profile. Consequently, we can define the average concentration of species $i$ as
\begin{equation}
c_{i}\left( t \right) = \frac{N_{i}}{V_{i}\left( t \right)},
\label{q3.2}
\end{equation}
where $N_{i}$ is the number of species $i$ and $V_{i}(t)$ is its time dependent average cluster volume at time $t$.  The total number of species $N_{i}$  does not change by influence of diffusion process. 
When we calculate the derivative of Eq. (\ref{q3.2})  we obtain  
\begin{equation}
  \frac{dc_{i}}{dt}=-\frac{c_{i}}{V_{i}}\frac{dV_{i}}{dt},
\label{q3.4}
\end{equation}
which expresses the influence of diffusion process only.

 We can substitute now the first term of the right-hand side of the equation (\ref{qB2}) that expresses the influence of diffusion  by the expession (\ref{q3.4}), which means that the diffusion of species i is expressed by corresponding time dependent volume $V_i$.  Consequently, it is possible to substitute the system of partial differential equations by the system of ordinary differential equations 
\begin{equation}
  \frac{dc_{i}}{dt}=-\frac{c_{i}}{V_{i}} \frac{dV_{i}}{dt} -c_{i} \sum_{j}k_{ij}c_{j} +  \sum_{j,k\neq i}k^{(i)}_{jk}c_{j}c_{k}.
\label{q3.6}
\end{equation}

It means that in the fully described system also the numbers $N_i$ decrease with time (see Eq. (\ref{q3.2}). And it is advantageous to substitute the time dependence of $c_{i}(t)$ by time dependence of $N_{i}(t)$; i.e., the concentration is substituted by the number of $i$ objects in radical cluster, which is also in agreement with the fact that the number of SSB formed in DNA should be proportional to the number of radicals in cluster. Substituting $c_{i}(t)$ by $N_{i}(t)$ with the help of Eq. (\ref{q3.2}) the Eq.(\ref{q3.6}) may be rewritten as
\begin{equation}
\frac{dN_{i}}{dt}=-N_{i}\sum_{i}k_{ij}\frac{N_{i}N_{j}}{V_{j}} + V_{i} \sum_{j,k\neq i}k^{(i)}_{jk}\frac{N_{j}N_{k}}{V_{j}V_{k}},
\label{eq22}
\end{equation}
which is again the system of ordinary differential equations.

The functions $N_i(t)$ may be determined from the last equation (\ref{eq22})  if the individual $V_i(t)$ functions may be regarded as given, which may be derived from Eq. (\ref{q1.37}). The volume $V_i(t)$ may be then given as 
\begin{equation}
   V_i(t) = \frac{4}{3}\pi\bar{r}_i^3(t)  \label{vol}
\end{equation}   
where the average radius of the i species cluster equals
\begin{equation}
   \bar{r}_{i}\left(t\right) = \frac{1}{N_i(t)}\int_0^{\infty} r\,
   c_{i}\left(r,t\right)4\pi r^{2}dr;
\label{q1.44}
\end{equation}
and inserting according to Eq. (\ref{q1.37}) one obtains
\begin{equation}
V_{i}\left( t \right) = \frac{256}{3} \sqrt{\left( \frac{D_{i}^{3}t^{3}}{\pi}\right)}.
\label{q1.49}
\end{equation}
  From Eqs. (\ref{eq22}) and (\ref{q1.49}) one obtains then the final system of ordinary differential equations
\begin{equation}
\frac{dN_{i}}{dt}=-\frac{3}{256}\sqrt{\left( \frac{\pi}{t^{3}}\right)}\left( N_{i}\sum_{j}k_{ij} \frac{N_{j}}{\sqrt{\left( D^{3}_{i} \right)}} + \sqrt{\left( D^{3}_{i}\right)}\sum_{j,k\neq i} k^{(i)}_{jk} \frac{N_{j}N_{k}}{\sqrt{\left( D^{3}_{j}D^{3}_{k}\right)}}\right).
\label{q3.7}
\end{equation}

\section*{\large\sf  Radical cluster and DSB formation}

Before passing to the analysis of experimental data we must distinguish between two kinds of species $i=1,..,n$ involved in chemical reactions. Some of them, i.e., $i=1,..,s<n$, are assumed to be radicals formed by the impact of ionizing particles, while the other are formed in the course of diffusion or are present permanently in the water solution. Thus the radical cluster is formed by the first kind of species that decrease quickly as the consequence of corresponding chemical reactions. 

It is evident that the actual origin of the cluster cannot be identified with $t = 0$ of cluster volume evolution described by Eq. (\ref{q1.37}).  It is necessary to introduce the time $t_0$ that corresponds to initial values $N_i(t_0), \;i=1,..,s,$ forming a radical cluster immediately after the impact of ionizing radiation; the radicals reacting then with DNA molecules and forming individual SSB. However, according to our earlier results \cite{lokaj2} an efficient cluster may be formed also in a certain distance from a DNA molecule; contact being realized later due to cluster diffusion and DNA molecule motion. Thus the probability of SSB formation by a radical $i$ will equal  
\begin{equation}
      p_{i}=\int^{t_{m}}_{t_{0}}\alpha_{i}N_{i}(t)dt,
\label{eq29}
\end{equation}
where parameters $\alpha_{i} \;(i = 1,\ldots ,s)$ must be proportional to the reaction rates of individual species $i$ with DNA molecules given in literature. As to the final time value it might be put $t_m\rightarrow\infty$; however, practically it is sufficient to integrate to $t_m$ corresponding to $N_i(t_m)<1$.  

And for the probability $p_{s}$ of SSB formation in a DNA molecule it holds then
\begin{equation}
   p_{s}=\sum_{i}p_{i}+p_{dir},
\label{eq28}
\end{equation}
where parameter $p_{dir}$ represents a direct effect of the given ionizing radiation that might be additional to the indirect chemical mechanism. However, we are interested in radical clusters  being able to form DSB in corresponding DNA molecules.  The corresponding probability may be then given by
\begin{equation}
        p_{D}=p_{S}^{2},
\label{eq30}
\end{equation}
which may be correlated to experimentally established values of DSB numbers under different conditions (oxygen content in our case).

\section*{\large\sf Specification of the model according to data kind}

As already mentioned we shall analyze the data concerning DSB formation under different oxygen content. And we shall assume that the corresponding average radical cluster at time $t_{0}$ consists of the following species:  
$H^{\bullet}$, $OH^{\bullet}$, $e^{-}_{aq}$.   %, $HO^{\bullet}_{2}$, $H_{3}O^{+}$, $O^{-}_{2}$. Oher radicals depending on oxygen presence are assumed to be formed a little later by reactions with oxygen molecules (depending on oxygen content).

The other species playing the role in final effect are then introduced in Table \ref{tab:3} together with chemical reactions being considered in the right-hand side of Eq. (\ref{q3.7}); the reaction rates (in corresponding units) established experimentally are also given. Only a part of reactions introduced in Table \ref{tab:1} have been taken into account. 

\begin{table}[htbp]
	\centering
  \small
		\begin{tabular}{|rrcl|c|}
			\hline
			&&&&  \textbf{Rate constants} \\
			\multicolumn{4}{|c|}{\textbf{Reactions}} & (nm$^{3}$. N$^{-1}$. ns$^{-1}$)\\
			\hline
			1. & $H^{\bullet}+H^{\bullet}$ & $\longrightarrow$ & $H_{2} $ & $16.606$ \\
			2. & $e^{-}_{aq} + H^{\bullet}$ & $\longrightarrow$ & $H_{2} + OH^{-}$ & $41.514$\\
			3. & $e^{-}_{aq} + e^{-}_{aq}$ & $\longrightarrow$ & $H_{2} + 2OH^{-}$& $9.963$\\
			4. & $e^{-}_{aq} + OH^{\bullet}$ & $\longrightarrow$ & $OH^{-} + H_{2}O$ & $49.817$\\
			5. & $H^{\bullet} + OH^{\bullet}$& $\longrightarrow$ & $H_{2}O$& $39.854$\\
			6. & $OH^{\bullet}+OH^{\bullet}$ & $\longrightarrow$ & $H_{2}O_{2}$ & $6.642$\\
			7. & $H_{3}O^{+} + e^{-}_{aq}$ & $\longrightarrow$ & $H^{\bullet} + H_{2}O $& $38.193$\\
			8. & $HO^{\bullet}_{2} + H^{\bullet}$ & $\longrightarrow$ & $H_{2}O_{2}$ & $16.606$\\
			9. & $e^{-}_{aq} + O_{2}$ & $\longrightarrow$ & $O^{-}_{2} + H_{2}O$ & $31.551$\\
			10. & $HO^{\bullet}_{2} + OH^{\bullet}$ & $\longrightarrow$ & $H_{2}O + O_{2}$ & $16.606$\\
			11. & $HO^{\bullet}_{2} + HO^{\bullet}_{2}$ & $\longrightarrow$ &  $H_{2}O_{2} + O_{2}$ & $0.003$\\
			12. & $H^{\bullet} + O_{2}$ & $\longrightarrow$ & $HO^{\bullet}_{2}$ & $16.606$\\
			13. & $O^{-}_{2} + H_{3}O^{+} $ & $\longrightarrow$ & $HO^{\bullet}_{2}$ & $49.817$\\
			\hline
		\end{tabular}
	\caption{Recombination reactions chosen from Table \ref{tab:1}} 
	\label{tab:3}
\end{table}

The corresponding system of ordinary differential equations, which describe dynamic of recombination and diffusion processes in the considered case may be then written in the form 
\begin{eqnarray}
\frac{dN_{H}}{dt}=-\frac{3N_{H}}{256}\sqrt{\left( \frac{\pi}{t^{3}}\right)}\cdot \left( \frac{2k_{1}N_{H}}{\sqrt{\left(D^{3}_{H}\right)}}  
+ \frac{k_{2}N_{e}}{\sqrt{\left(D^{3}_{e}\right)}}
+\frac{k_{5}N_{OH}}{\sqrt{\left(D^{3}_{OH}\right)}}
+\frac{k_{8}N_{O}}{\sqrt{\left(D^{3}_{O}\right)}}\right)+
\nonumber \\
+\frac{3}{256}\sqrt{\left(\frac{\pi}{t^{3}} \right)}\cdot\sqrt{\left( D^{3}_{H}\right)}
%\cdot \nonumber \\ 
\cdot   
\frac{k_{7}N_{H_{3}O}N_{e}}{\sqrt{\left(D_{H_{3}O} D_{e}\right)^{3}}}
%+\frac{k_{12}N_{OH}N_{H_{2}}}{\sqrt{\left(D_{OH}\cdot D_{H_{2}}\right)^{3}}}
- N_{H}\left( k_{12}\left[O_{2}\right]+k_{15}\right)
\label{qA}
\end{eqnarray}

\begin{eqnarray}
\frac{dN_{OH}}{dt}=-\frac{3N_{OH}}{256}\sqrt{\left( \frac{\pi}{t^{3}}\right)}\cdot \left( 
\frac{k_{5}N_{H}}{\sqrt{\left(D^{3}_{H}\right)}}
+\frac{2k_{6}N_{OH}}{\sqrt{\left(D^{3}_{OH}\right)}}
+\frac{k_{10}N_{O}}{\sqrt{\left(D^{3}_{O}\right)}}
+\frac{k_{4}N_{e}}{\sqrt{\left(D^{3}_{e}\right)}}	\right)
- \nonumber \\
- N_{OH}\left( k_{14}\left[O_{2}\right]+k_{16}\right)
\label{qC}
\end{eqnarray}

\begin{eqnarray}
\frac{dN_{e}}{dt}=-\frac{3N_{e}}{256}\sqrt{\left( \frac{\pi}{t^{3}}\right)}\cdot \left( \frac{k_{2}N_{H}}{\sqrt{\left(D^{3}_{H}\right)}}  
+\frac{k_{4}N_{OH}}{\sqrt{\left(D^{3}_{OH}\right)}}
+ \frac{2k_{3}N_{e}}{\sqrt{\left(D^{3}_{e}\right)}}
+\frac{k_{7}N_{H_{3}O}}{\sqrt{\left(D^{3}_{H_{3}O}\right)}}
\right)-
\nonumber \\
-k_{9}N_{e}\left[O_{2}\right]
\label{qB}
\end{eqnarray}

\begin{eqnarray}
\frac{dN_{O}}{dt}=-\frac{3N_{O}}{256}\sqrt{\left( \frac{\pi}{t^{3}}\right)}\cdot \left(\frac{k_{8}N_{H}}{\sqrt{\left(D^{3}_{H}\right)}}  
+\frac{k_{10}N_{OH}}{\sqrt{\left(D^{3}_{OH}\right)}}  
+\frac{2k_{11}N_{O}}{\sqrt{\left(D^{3}_{O}\right)}} \right)+ \;\;\;\;\;
\nonumber \\
+\frac{3}{256}\sqrt{\left(\frac{\pi}{t^{3}} \right)}
\cdot\sqrt{\left( D^{3}_{O}\right)}
\cdot \frac{k_{13}N_{H_{3}O}N_{O^{-}_{2}}}{\sqrt{\left(D_{H_{3}O}D_{O^{-}_{2}}\right)^{3}}}
+ k_{12} N_{H}\left[O_{2}\right]-k_{17}N_{O}
\label{qH}
\end{eqnarray}

\begin{eqnarray}
\frac{dN_{H_{3}O}}{dt}=-\frac{3N_{H_{3}O}}{256}\sqrt{\left( \frac{\pi}{t^{3}}\right)}\cdot \left( 
\frac{k_{7}N_{e}}{\sqrt{\left(D^{3}_{e}\right)}}
+\frac{k_{13}N_{O_{2}^{-}}}{\sqrt{\left(D^{3}_{O_{2}^{-}}\right)}}\right)\label{qD}
\end{eqnarray}

\begin{eqnarray}
\frac{dN_{O^{-}_{2}}}{dt}=-\frac{3N_{O^{-}_{2}}}{256}\sqrt{\left( \frac{\pi}{t^{3}}\right)}\cdot 
\frac{k_{13}N_{H_{3}O}}{\sqrt{\left(D_{H_{3}O}^{3}\right)}}+k_{9}N_{e}\left[0_{2}\right] \;\;\;
\label{qJ}
\end{eqnarray}
where $k_{1},\ldots , k_{13}$ are rate constant of reactions given in Table \ref{tab:3} and $k_{14},\ldots , k_{17}$ correspond to additional reactions the rate constants of which were not available; they are taken as fully free parameters in the optimization procedure. On the other side the values of involved diffusion coefficients have been fixed as already mentioned (see Table \ref{tab:2}).

In agreement with general opinion we will assume, that  for SSB formation practically the radicals $H^{\bullet}, OH^{\bullet}, e^{-}_{aq}$ and $HO^{\bullet}_{2}$ are responsible. The probability $p_{s}$ of SSB formation in DNA molecules depends then on the instant of encounter of a molecule with a given cluster. One can write for an average cluster
\begin{eqnarray}
p_{S}=p_{H}+p_{OH}+p_{e}+p_{O}+p_{1},
\label{A5}
\end{eqnarray}
where the individual probabilities are given as averages over different $t$:

\begin{eqnarray}
p_{H}=\int^{t_{m}}_{t_{0}}\alpha_{H}N_{H}\left(t\right)dt
\label{A1}
\end{eqnarray}

\begin{eqnarray}
p_{OH}=\int^{t_{m}}_{t_{0}}\alpha_{OH}N_{OH}\left(t\right)dt
\label{A2}
\end{eqnarray}

\begin{eqnarray}
p_{e}=\int^{t_{m}}_{t_{0}}\alpha_{e}N_{e}\left(t\right)dt,
\label{A4}
\end{eqnarray}

\begin{eqnarray}
p_{O}=\int^{t_{m}}_{t_{0}}\alpha_{O}N_{O}\left(t\right)dt
\label{A3}
\end{eqnarray}
and parameters $\alpha_{H}, \;\alpha_{OH}, \;\alpha_{O},\;\alpha_{e}$ are held to be proportional to the reaction rates of individual radicals and of  $e^{-}_{aq}$  with DNA molecules (see, e.g., \cite{pika}). The value $t_{m}$ should correspond to cluster diffusion time (e. g. to time when less than one radical of the given type is present in a given cluster). The parameter $p_{1}$ in Eq. (\ref{A5}) represents direct effect under the given irradiation conditions.

The probability of DSB formation may be then given by
\begin{eqnarray}
p_{D}=p_{S}^{2}.
\label{A6}
\end{eqnarray}

\section*{\large\sf Analysis of experimental data}

As mentioned we have used already the preliminary version of the given mathematical model in analyzing experimental data of Blok and Loman \cite{blok}. Even if their paper was published in 1973 the data represent still very important information about the formation of DSB (and also SSB) at different oxygen concentrations.  The data were gained by irradiating the $\Phi$X174-DNA in water solution by photons ($\sim$1,25 MeV) of Co-60 isotope; the applied dose was 5 Gy. The  solution contained  25 $\mu g.ml^{-1}$ of DNA in 0,01 phosphate buffer at pH7. The main measurements concerned SSB numbers at different oxygen concentrations while the corresponding DSB numbers were established in some cases only to determine the ratio of DSB and SSB. We have established the corresponded dependence of DSB numbers on oxygen concentrations from the ratio of DSB and SSB numbers in all measured points (see Fig. 1). We have been interested in DSB formation mainly as they are responsible for  biological effect of ionizing radiation. While the experimental data have contained the results for two different gas mixtures, in the following the model will be demonstrated being applied only to the mixture $N_2-O_2$. 
 
\begin{figure}[ht]
	\centering
		\includegraphics[width=12cm]{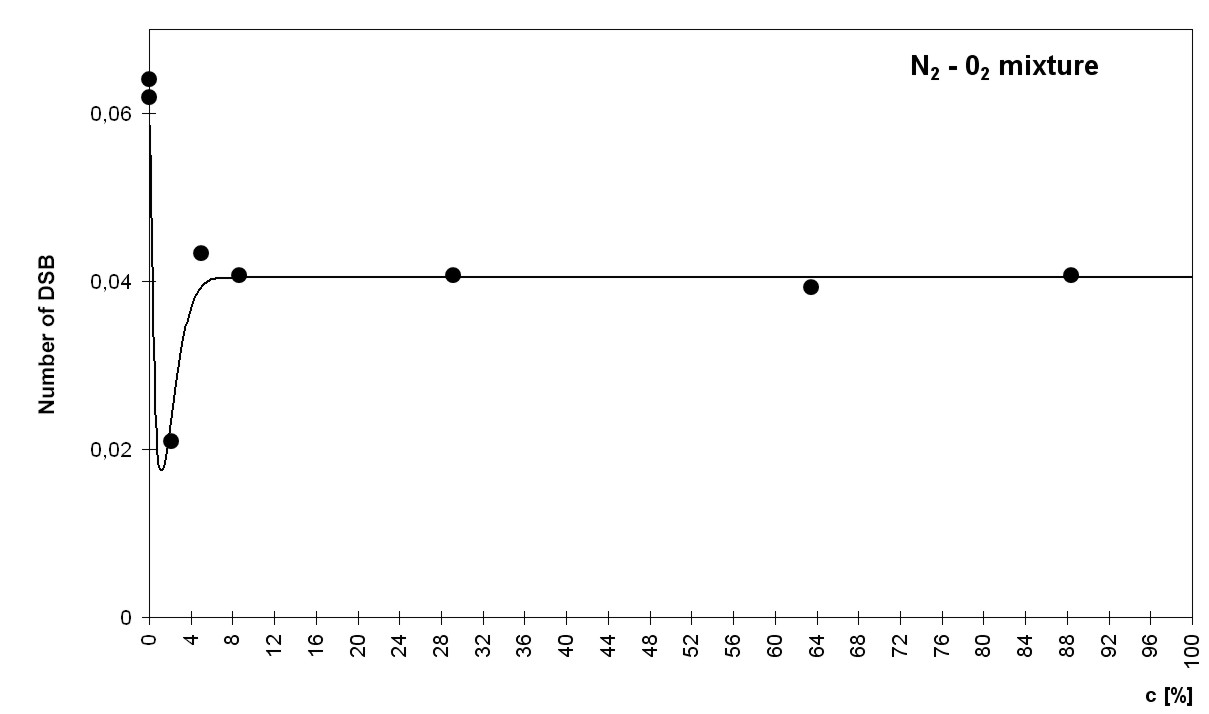}
	\caption{DSB numbers (per DNA molecule) at different oxygen concentrations in $N_2-O_2$ mixture; experimental data - individual points, theoretical dependence - full line. }
	\label{fig:1}
\end{figure}

We have assumed that the DSB may be created in the case only when a greater radical cluster (formed during irradiation) meets a DNA molecule. And we have been looking for what the size of average cluster may be and which chemical processes may be running inside such a cluster before it diffuses. We have required for all chemical reactions being involved to possess chemical reaction rates corresponding approximately to values (or rather to their mutual ratios) being introduced in literature. And we have asked also for the ratios of initial numbers or radicals  $N_{H^\cdot}$,  $N_{OH^\cdot}$, $e^-_{aq}$ to correspond to frequencies of processes leading to their formation, i.e.,
  \[   N_{H^\cdot} \;:\;  N_{OH^\cdot} \;:\; N_e  \;\;\;\cong\;\;\; 1 \;:\; 2 \;:\; 2 \:.   \] 
The initial values of other radicals have been put to be zero at time $t_0$ as they may be formed with some delay after the original water radicals meet an $O_2$ molecule. The agreement with literature data (with their mutual ratios) was asked also for reaction rates of different radicals with DNA molecules; i.e., for the ratios of quantities $\alpha_H, \;\alpha_{OH}, \;\alpha_e , \;\alpha_O$. Only the rest of parameters have been left free to be determined in optimization.

The actual values of individual parameters have been determined with the help of the MINUIT program (see \cite{james}). The best fit to the given experimental data was obtained on the basis of the following values: For the initial numbers of radicals forming average cluster at $t_0$ it has been obtained 
     \[   N_{H^\cdot}  =  18.11,  \;\;  N_{OH^\cdot}  =  31.9, \;\; N_e  =  35.0.   \]
And for the reaction rates of different radicals with DNA (comp. \cite{pika})
    \[  \alpha_H \;:\; \alpha_{OH} \;:\;\alpha_e\;:\; \alpha_O  \;\; = \;\; 5.95  : 93.53  : 0.01 : 0.58.  \]
The probability of a direct effect has been approximately
     \[    p_1  =   0.0003  \; ,     \]
indicating to be practically negligible. 
The values of parameters $k_1,...,k_{13}$ have been introduced in Table 3.
And for the additional reaction rates it has been obtained 
  \[    k_{14}  =   325.54 , \;\; k_{15}  =  0.098 , \;\; k_{16}  =   0.107 , \;\; k_{17}  =   0.004 .\]
The model dependence is shown in Fig. 1  together with the corresponding experimental data.

Experimental curve exhibits, however, a quite flat dependence for higher values of oxygen concentration. It follows from the fact that the oxygen does not dissolve fully in water at all concentrations; the saturation depending on concentration value. It is then necessary to write for oxygen concentration $\left[O_2\right]$ in water solution (see Eqs. (\ref{qA}-\ref{qJ}))
\begin{equation}
    \left[O_2\right] \;=\; A_O \{1 - \exp(-(c\sigma_O)^{\tau_O})\}
\end{equation}  
where $c$ is the oxygen concentration in gas mixture. Our fit has been then obtained when this dependence has been added with the following values of the involved parameters 
    \[ A_O \,=\, 0.0004, \;\; \sigma_O \,=\, 43.68, \;\; \tau_O \,=\, 1.7;  \]
the value of parameter $A_O$ representing maximal solubility in the given solution seems to be in good agreement with the value $0.0008\,nm^{-3}$ for pure water.      

\section*{\large\sf Conclusion}

The described model enables to study the influence of corresponding substances and of individual processes running during the chemical stage on the important damages of DNA molecules when irradiated by ionizing radiation. These processes may play important role especially for low-LET radiation (at not very high doses and dose rates) when individual DSB are formed mainly by track ends of secondary electrons. The approach might be easily adapted also to track ends of protons (or similarly) if the initial form of radical clusters is assumed to be cylindrical (instead of spherical) and the influence of diffusion is modified in corresponding way.

In the presented paper we have demonstrated the influence of different oxygen content on DSB formation and the fact that the rather drastic experimental dependence in the region of very small oxygen content may be reproduced when all chemical reaction and diffusion rates are in agreement with ratios between corresponding values established in other experiments. Here we have tried to describe mainly the corresponding mathematical model and its solution approach without going to conclusions concerning proper radiobiological aspects. These goals will be followed in another paper. 

The model has opened, however, some other possibility of analyzing the influence of different contents of other present substances that may act also as radiomodifiers, enlarging or diminishing the radiobiological effect. When the reaction rates between individual substances are known their effect at their divers concentrations and also at different oxygen concentrations might be established. A series of corresponding experimental data that might be analyzed with the help of the presented model may be found in literature.    
\\

%\newpage
%\vspace{0.5cm}

\end{document}